\documentclass[conference,compsoc]{IEEEtran}

\ifCLASSOPTIONcompsoc
  \usepackage[nocompress]{cite}
\else
  \usepackage{cite}
\fi
\ifCLASSINFOpdf
  \usepackage[pdftex]{graphicx}
\else
  \usepackage[dvipdfmx]{graphicx}
\fi
\begin{document}

\title{Evaluation of the RIKEN Post-K Processor Simulator}

\author{\IEEEauthorblockN{Yuetsu Kodama\IEEEauthorrefmark{1},
    Tetsuya Odajima\IEEEauthorrefmark{1},
    Akira Asato\IEEEauthorrefmark{2} and
    Mitsuhisa Sato\IEEEauthorrefmark{1}}
\IEEEauthorblockA{\IEEEauthorrefmark{1}Center for Computational Science(R-CCS)\\
  Email: \{yuetsu.kodama,tetsuya.odajima,msato\}@riken.jp}
\IEEEauthorblockA{\IEEEauthorrefmark{2}Fujitsu Limited\\
  Email: asato@jp.fujitsu.com}
}

\maketitle

\begin{abstract}

For the purpose of developing applications for Post-K at an early
stage, RIKEN has developed a post-K processor simulator. This
simulator is based on the general-purpose processor simulator gem5.
It does not simulate the actual hardware of a post-K
processor. However, we believe that sufficient simulation accuracy can
be obtained since it simulates the instruction pipeline of out-of-order
execution with cycle-level accuracy along with performing detailed parameter
tuning of out-of-order resources and function expansion of cache/memory
hierarchy. In this simulator, we aim to estimate the execution cycles of
one node application on a post-K processor with accuracy that enables
relative evaluation and application tuning. In this
paper, we show the details of the implementation of this simulator and
verify its accuracy compared with that of a post-K test chip.

\end{abstract}

\section{Introduction}

RIKEN is developing a post-K computer as a next-generation flagship
computer for Japan. Post-K is planned to start operation around 2021,
but there is still more than a year to provide early access
whereby users can use preliminarily a part of the actual machines. RIKEN
has developed a RIKEN post-K processor simulator with the aim of
developing applications on the Post-K as soon as possible.

The RIKEN simulator is based on the general-purpose processor
simulator gem5~\cite{gem5}. In gem5, out-of-order execution based on
the basic architecture model can be simulated and the correct number
of execution cycles can be estimated. This basic architecture model
differs in certain details from the Fujitsu A64FX
processor~\cite{A64FX-hotchip, A64FX-cluster} which is the processor
of the Post-K. However, we believe that sufficient accuracy
can be obtained since it simulates the 
out-of-order pipeline with cycle-level accuracy along with performing
detailed parameter tuning of out-of-order resources and function
expansion of the cache/memory hierarchy. In this simulator, we aim to
estimate the execution cycles of one node application on a post-K
processor with accuracy that enables relative evaluation and 
application tuning.

In this paper, we first show an overview of the A64FX which is going to be the
Post-K processor, and explain the general purpose processor simulator,
gem5, as the basis of our simulator. Next, we show the details of the
implementation of the RIKEN simulator. Then we compare the execution
time of this simulator with the execution time of the A64FX test chip,
and verify its accuracy. Finally we summarize this paper.

\begin{figure*}[!t]
  \begin{center}
    \includegraphics[width=12cm]{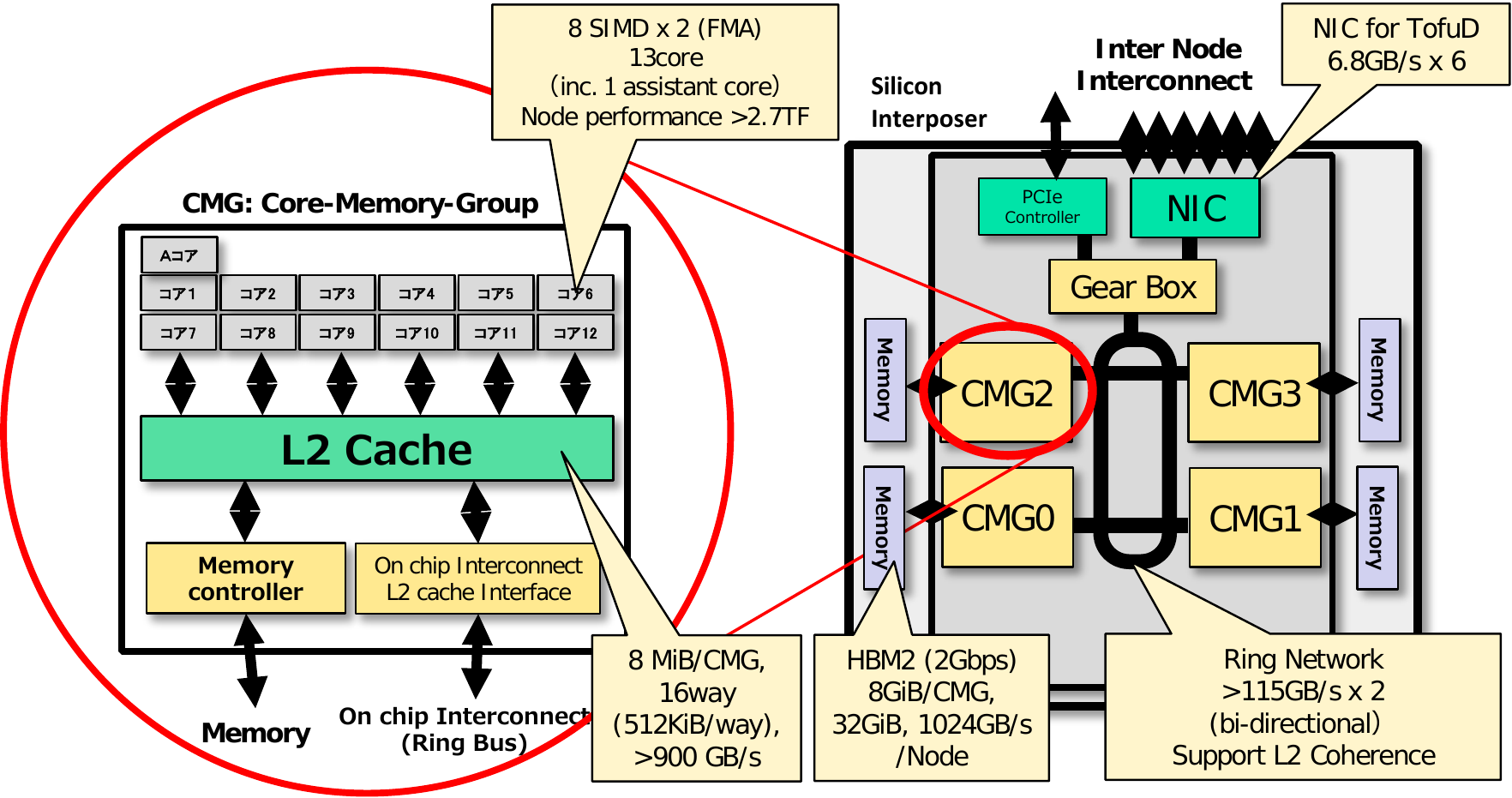}
    \caption{Configuration of the Post-K processor, A64FX}\label{fig:A64FX}
  \end{center}
\end{figure*}

\section{Post-K processor, the A64FX}

Figure \ref{fig:A64FX} shows the configuration of the post-K
processor, the A64FX, developed by Fujitsu. The A64FX consists of four CMGs (Core Memory Group) connected via a
ring bus. Each CMG has 13 cores (including an OS support core
called an assistant core), a shared L2 cache and a memory
controller. The L2 cache has a capacity of 8 MiB per CMG, has a
throughput exceeding 900 GB/s, and supports coherence between
CMGs on a chip. The memory HBM2 is mounted in the same package with a capacity of
8 GiB per CMG and has a throughput of 256 GB/s per CMG.

Each core is based on the 64-bit architecture model of the Armv8.2-A and supports a new
SIMD extension called SVE (Scalable Vector Extension)~\cite{SVE}. Each
core has two SIMD pipelines of 512 bits, and the computing performance
of the entire processor exceeds 2.7 TFLOPS. In addition to supporting
double precision and single precision floating point numbers, SVE
supports half precision and also supports 64 bit, 32 bit, 16 bit, and 8
bit integers. Each core has a 64 KiB data and instruction L1
cache. The data cache load performance exceeds 230 GB/s and the
store performance exceeds 115 GB/s.

The A64FX has a network interface in the processor that supports the
6-dimensional torus network (TofuD), the same as the K Computer. Each link
speed has been expanded to 6.8 GB/s, and the number of transfer engines has been
increased to six.

\section{Processor Simulator gem5}

The RIKEN simulator is based on the open source processor simulator
gem5. The main features of gem5 are as follows. For details, please
refer to http://gem5.org.

\begin{itemize}

\item It supports multiple instruction set architectures (ISA), such as
  Alpha, Arm, SPARC, x86, RISC-V, and GPU, etc. The RIKEN simulator only
  covers the Armv8-A architecture.

\item It supports two system modes. One is full system (FS) mode which
  simulates the OS code with the processor on the simulator. The other
  is system emulation (SE) mode which simulates the system call by
  software. The RIKEN simulator supports SE mode only.

\item It supports multiple CPU models. The Atomic model simulates a CPU
  with 1 cycle execution of one instruction. The Timing model simulates a
  CPU with the timing of a memory reference added to the Atomic model. The InOrder
  model simulates a CPU with an in-order pipeline. The Out-of-order (O3)
  model simulates a CPU with an out-of-order pipeline. In the RIKEN
  simulator, only the Atomic model and the O3 model are supported.

\item It has two memory systems. One is the simple Classic model and the
  other is the Ruby model that can flexibly configure a coherent memory
  system. Currently, in the RIKEN simulator, the Classic model is used.

\end{itemize}

\section{RIKEN simulator}

Gem5 supports the Armv8-A ISA, but in 2016 when we began developing the
RIKEN simulator, SVE (Scalable Vector Extension), a new SIMD
extension, was supported only in Atomic mode, not in O3
mode. Therefore, we decided to independently develop an O3 mode for
SVE. We are offering the implementation of SVE in Atomic mode from Arm
Research which maintains the Armv8-A ISA, and we have also developed our own
O3 mode for SVE.

After that, Arm Research also developed an O3 mode for SVE. We
compared both implementations, and found that there was no big
difference. Since the Arm version will be integrated into the main gem5
distribution, we decided to move to the SVE O3 implementation of the Arm
version. The transition began in April 2018, and was
completed in October 2018.

In the O3 mode of gem5, it is possible to flexibly specify detailed
parameters such as the latency and the throughput of each stage of the
pipeline, as well as each resource size for out-of-order execution. We received
the detailed parameters of the A64FX processor from Fujitsu. By
adjusting these parameters, we can simulate programs with sufficient
accuracy within the processor. However, since the CPU model of gem5
is based on an out-of-order pipeline based on that of the Alpha 21264, and
fundamentally differs from the pipeline configuration of the A64FX, the
difference exists. The main differences are as follows.

\begin{itemize}

\item While there is one reservation station in gem5, the A64FX is
  divided into a total of four: one for memory address calculation, two
  for arithmetic operations, and one for branch operations. The RIKEN simulator makes
  adjustments with values close to the four total values, so if processing
  units are used that are biased towards any of them, the difference between the two
  may become large.

\item In gem5, the memory address calculation is performed in the
  load/store unit, so the store instruction is executed after both the
  memory address operand and the write data operand are determined. On
  the other hand, in the A64FX, memory addresses can be calculated in
  independent units. Since the RIKEN simulator is the same as gem5, there
  may be differences in the timing of address calculation.

\item In gem5, it is assumed that one instruction uses a single
  execution unit, while in the A64FX, some instructions may use multiple
  execution units in parallel or in sequential form. In the RIKEN
  simulator, one instruction is assigned to one main execution unit,
  assuming that the frequency of such an instruction is low. Also,
  some instructions, such as gather load instructions, are divided
  into multiple micro instructions and executed using the micro
  instruction mechanism of gem5. Such execution will use more
  out-of-order resources than in the A64FX and may appear as a difference in
  execution time.

\end{itemize}

The RIKEN simulator currently supports only one CMG simulation, and
multi-thread execution of up to 12 cores is possible. In order to
accurately estimate the execution time of a program, it is necessary
to accurately simulate not only the instruction execution time, but
also the access time of the cache memory hierarchy. Therefore, in the
RIKEN simulator, we expanded gem5 to match the performance of the cache
memory hierarchy to that of the A64FX. The main extensions are as follows.

\begin{itemize}

\item The L1 cache and L2 cache capacity, associativity, line size and
  latency were set according to the actual settings of the A64FX by setting the gem5
  parameters.

\item In gem5, it is assumed that load and store can simultaneously
  access the L1 cache in the same cycle. On the other hand, the A64FX enables
  two load or single store operations. The RIKEN simulator enabled the same
  controls as those of the A64FX.

\item In gem5, cache fill from L2 cache to L1 cache is controlled independently of
  access to the L1 cache of the core. Therefore, the performance of the L1 cache will be
  enhanced. By performing exclusive control between these, the RIKEN
  simulator was able to simulate the L1 cache performance of the A64FX
  accurately.

\item In gem5, when access to the L1 cache exceeds cache alignment, overhead is
  generated because the access is divided into multiple
  accesses. The A64FX is designed so as not to cause performance degradation
  even when accessing across cache lines. The RIKEN simulator also added
  such a function.

\item In gem5, software prefetch was supported, but only prefetch for read
  access was implemented. Prefetch for store access is also
  important for optimizing memory access, and thus was added in the RIKEN
  simulator. This feature has already been reported to the gem5 developers
  and accepted.

\item Gem5 also has hardware prefetching capabilities, but only simple
  prefetching is supported. The RIKEN simulator added a K Computer
  compatible hardware prefetch function~\cite{K-computer}.

\item With gem5, you can easily specify that the L2 cache is to be shared with multiple
  cores, but since default the L2 cache is a single bank, L2 access is
  likely to be a bottleneck if the number of cores increases. Even
  with gem5, the L2 cache is designed as a module, so it is possible for
  users to set multiple banks of L2 cache by adding descriptions
  individually. The RIKEN simulator has been expanded so that the
  number of banks can be changed easily, simply by specifying
  parameters.

\item In gem5, the bus width between the L1 cache and the L2 cache is one
  parameter, and it is assumed that the transfer throughput from L1 cache
  to L2 cache and that from L2 cache to L1 cache are the same. On the other hand, in the
  A64FX, these two transfer throughputs are different. The RIKEN
  simulator has been extended to specify these two bus widths as
  different parameters. The same applies to the bus width between the L2
  cache and memory.

\item Gem5 supports HBM1, but HBM2 is not yet supported. The RIKEN
  simulator added HBM2 parameters based on the HBM1 parameters and set
  the bank access scheduling policy appropriately. Although the
  standard features of gem5 could not completely match the memory
  interleaving method used with the A64FX, the RIKEN simulator achieved
  almost the same memory performance as that of the A64FX by combining this feature with other
  parameters, such as burst length.

\end{itemize}

In addition to these, the following function enhancements have been
made to improve usability.

\begin{itemize}

\item Gem5 outputs various statistical information of the simulation to a
  file called stats.txt, which collects information from the start to
  the end of the simulation. However, since it is useful to obtain the
  statistical information on a specific section, a function to achieve this has
  been developed. However, in order to separate such information from stats.txt, it is necessary
  to specify the switching time at simulator startup. Therefore, 
  first, pre-execution is performed in order to obtain the time of the
  switching timing, and then, the time is specified to perform the
  main execution. We created a python script to do that.

\item The Fujitsu super computer system can acquire detailed profile
  information called PA data. We extended the statistical infromation
  in gem5 so that similar information could be acquired. In particular, we
  added a function to classify the cause of the wait instruction at the time of a 0 instruction
  commit into memory wait, an arithmetic operation wait,
  etc. Furthermore, a python script was created to extract information
  from stats.txt based on Fujitsu's PA data.

\item We extended the function used to obtain the cycle-by-cycle utilization
  of out-of-order resources.

\item We extended the function used to count element numbers of operations by an SIMD
  instruction, etc. Furthermore, a function to count only the number
  of valid operations according to the value of the predicate register
  has been added.

\item In gem5, for each 'OpClass' instruction group, which execution unit
  to use is specified, e.g., the instruction latency,
  etc. Normally, OpClass is determined for each opcode field of the instruction,
  but the A64FX has an instruction with different instruction latency,
  depending on the operand type (double precision or single
  precision). We have extended this item to accommodate such instructions.

\end{itemize}

We are considering whether to release these extended functions separately from
various detailed parameters involving NDA information from Fujitsu.

\section{Evaluation}

The RIKEN simulator was evaluated using several programs to find out
how accurate it is with the execution time of the A64FX. The evaluation
target is the test chip of the A64FX prototype, and does not indicate the
performance of the final post-K processor. Also, the compiler used for
generating the program to be executed was a prototype version of the
compiler for the Post-K computer from Fujitsu, which is the October 2018
version. It is the same version of the compiler used with the test chip
evaluation.

\subsection{Evaluation of the kernel program on a single core}

First, we compared the execution times of various kernel programs on a
single core with that of the test chip.

\begin{figure}[!t]
  \begin{center}
    \includegraphics[width=7cm]{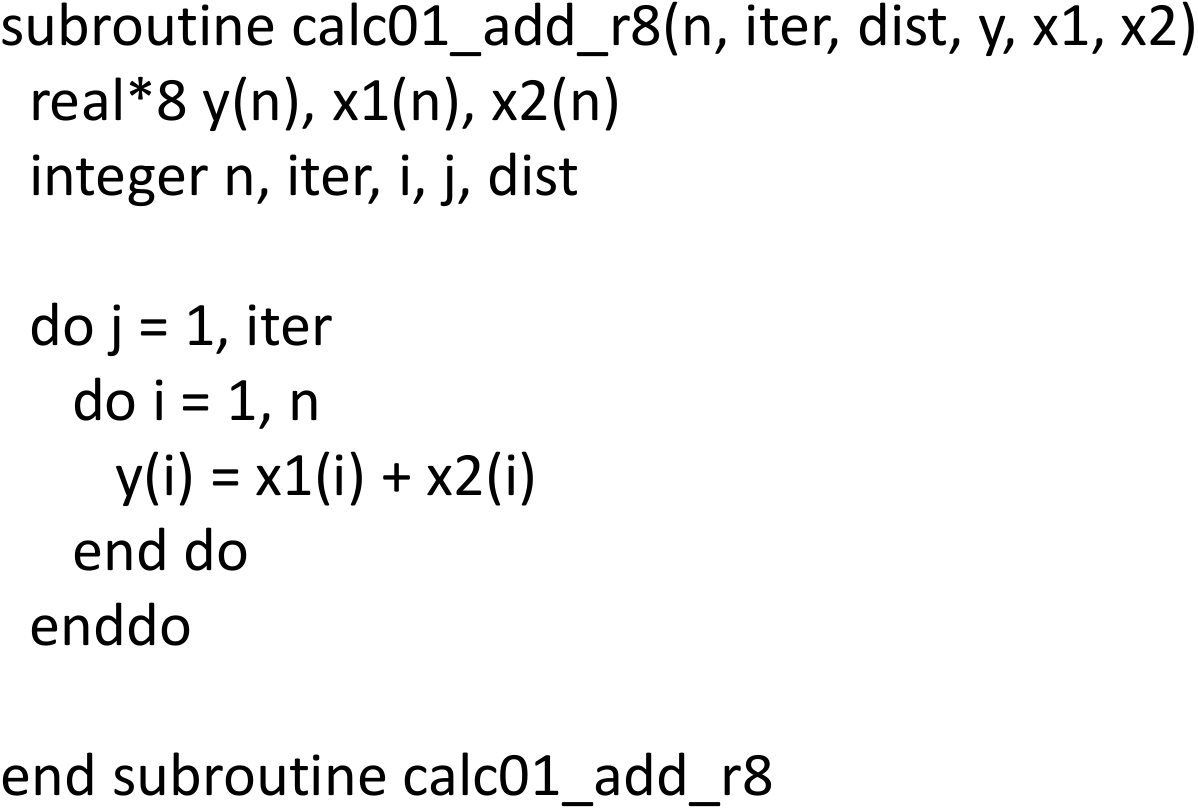}
    \caption{Example source code of the kernel program (double precision addition)}\label{fig:kernel}
  \end{center}
\end{figure}

\begin{table}[!t]
  \begin{center}
    \caption{Kernel program list}\label{tbl:kernel}
    \begin{tabular}{| l | l | r | l |} \hline
      Name & Type & Size & Kernel statement \\ \hline
      \multicolumn{4}{| l |}{Basic Arithmetic} \\ \hline
      add & addition & 2048 & y(i) = x1(i) + x2(i) \\
      sub & subtraction & 2048 & y(i) = x1(i) - x2(i) \\
      mul & multiplication & 2048 & y(i) = x1(i) * x2(i) \\
      fma & sum of products & 3072 & y(i) = y(i) + c0 * x1(i) \\
      div & division & 2048 & y(i) = x1(i) / x2(i) \\
      rev & reciprocal & 3072 & y(i) = 1 / x1(i) \\
      sqrt & square root & 3072 & y(i) = sqrt(x1(i)) \\ \hline
      \multicolumn{4}{| l |}{Type conversion} \\ \hline
      f2d & float to double & 4096 & y\_r8(i) = dble(x1\_r4(i)) \\
      i2d & integer to double & 4096 & y\_r8(i) = dble(x1\_i4(i)) \\
      d2f & double to float & 4096 & y\_r4(i) = real(x1\_r8(i)) \\
      d2i & double to integer & 4096 & y\_i4(i) = int(x1\_r8(i)) \\
      aint & aint conversion & 3072 & y\_r8(i) = aint(x1\_r8(i)) \\
      nint & nint conversion & 4096 & y\_i4(i) = nint(x1\_r8(i)) \\
      anint & anint conversion & 3072 & y\_r8(i) = anint(x1\_r8(i)) \\ \hline
      \multicolumn{4}{| l |}{Numeric function} \\ \hline
      abs & absolute value & 3072 & y(i) = abs(x1(i)) \\
      max & maximum value & 2048 & y(i) = max(x1(i), x2(i)) \\
      min & minimum value & 2048 & y(i) = min(x1(i), x2(i)) \\
      mod & remainder & 2048 & y(i) = mod(x1(i), x2(i)) \\
      sign & sign & 2048 & y(i) = sign(x1(i), x2(i)) \\ \hline
      \multicolumn{4}{| l |}{Mathematical function} \\ \hline
      atan & atan & 3072 & y(i) = atan(x1(i)) \\
      atan2 & atan2 & 2048 & y(i) = atan2(x1(i), x2(i)) \\
      cos & cos & 3072 & y(i) = cos(x1(i)) \\
      sin & sin & 3072 & y(i) = sin(x1(i)) \\
      exp & exp & 3072 & y(i) = exp(x1(i)) \\
      exp10 & exp10 & 3072 & y(i) = exp10(x1(i)) \\
      log & log & 3072 & y(i) = log(x1(i)) \\
      log10 & log10 & 3072 & y(i) = log10(x1(i)) \\
      pwr & power & 2048 & y(i) = x1(i) ** x2(i) \\ \hline
    \end{tabular}
  \end{center}
\end{table}

\begin{figure*}[!t]
  \begin{center}
  \includegraphics[width=12cm]{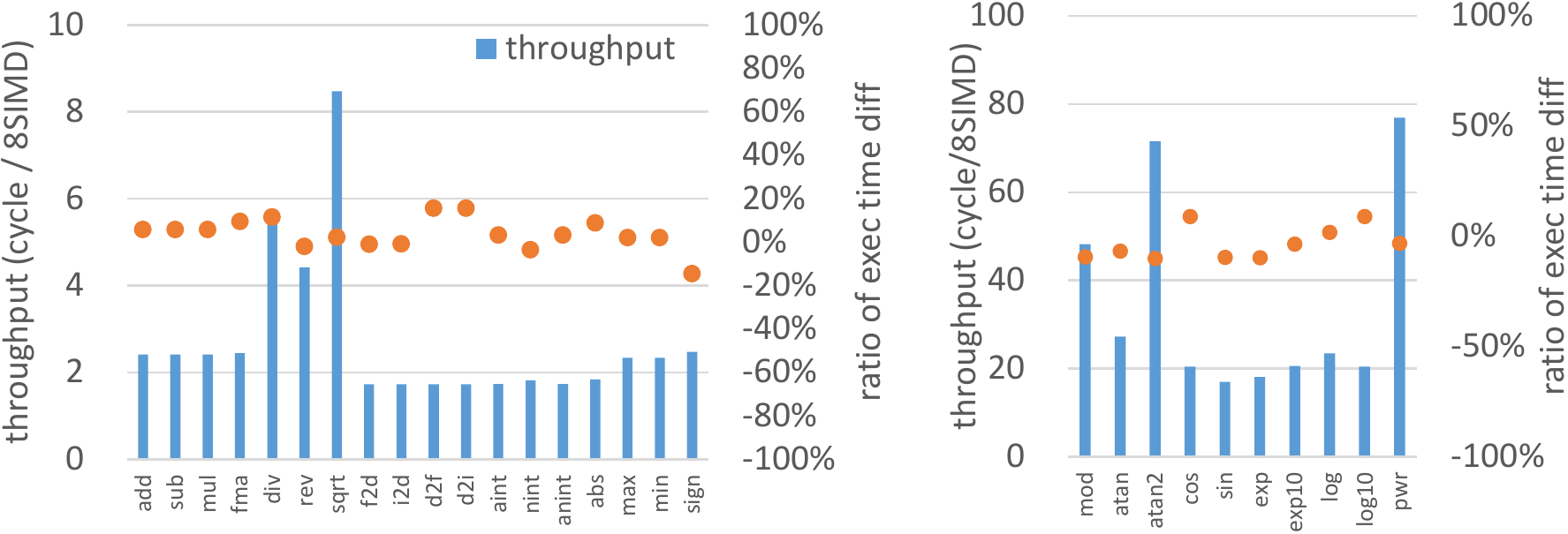}
  \caption{Execution throughput of each kernel in the RIKEN simulator and the execution time difference ratio of the test chip}\label{fig:kernelres}
  \end{center}
\end{figure*}

An example of a kernel program used for evaluation is shown in Figure
\ref{fig:kernel}, and a list of kernel programs used for evaluation is
shown in Table \ref{tbl:kernel}. There are four types of kernels:
basic arithmetic functions, type conversions, numerical functions, and
mathematical functions. The basic arithmetic functions include seven
double arithmetic operations: addition, subtraction, multiplication,
product-sum, division, reciprocal, square root. The type conversions
include seven conversions: conversion from double precision to single
precision and 32-bit integer and its inverse conversion, and
conversion from the double precision of 'aint', 'nint', and 'anint' that
are built-in functions of Fujitsu Fortran. The numerical functions
include five functions of double precision absolute value, maximum,
minimum, remainder and sign. The mathematical functions include nine
functions of double precision 'atan', 'atan2', 'cos', 'sin', 'exp',
'exp10', 'log', 'log10', and 'power'. Evaluation was performed on a
total of 28 kernels.

The column labeled 'Size' in Table \ref{tbl:kernel} represents the size of
each array referenced in the kernel part, and the size referring to
3/4 of the L1 cache as a whole is selected. It corresponds to the
number 'n' of iterations of the innermost loop. The number of iterations
of the outer loop, 'iter', is 1,000,000 for the test chip, but the
execution speed of the RIKEN simulator is about 10,000 times slower
than that of the actual machine, so the value of the outer loop is
1000 times of 1/1000 for the RIKEN simulator. The timer accuracy of
the RIKEN simulator, which is a cycle level simulator, is very high,
and even with a small number of iterations, it enables measurement of time
with sufficient accuracy.

These kernel programs were compared using the binary compiled with the
'-Kfast' option with the time executed on the test chip and the time
executed on the RIKEN simulator. Basically, all of these kernels are
executed by 8 SIMD in the Fujitsu compiler, and mathematical functions are
inlined, and optimization by software pipeline is effective. Divisions
and reciprocals are also calculated using reciprocal instructions that
are pipelined rather than using non-pipelined division instructions.

The evaluation results are shown in Figure \ref{fig:kernelres}. The bar
graph represents the operation throughput (the number of cycles
required for 8-element operation) evaluated by the RIKEN simulator
results, and corresponds to the left axis. The orange point is the
ratio of the difference in execution time between the test chip result and
the RIKEN simulator result, and corresponds to the right axis. An
execution time difference of 10\% indicates that the execution time of
the RIKEN simulator is 10\% longer than that of the test chip, and
-10\% indicates that the execution time of the test chip is 10\%
longer than that of the RIKEN simulator. Of the 28 kernels, only 5 
(div, d2f, d2i, sign, and atan2) showed a execution time difference
exceeding 10\%, and it was confirmed that the execution time difference
was 10\% or less with 80\% or more of the kernels. The average of
28 execution time differences is 1.3\%, the standard deviation is 7.8\%,
and the average of absolute values is 6.6\%.

Details of the causes of the execution time differences are under
consideration. In d2f and d2i, the merge effect of the write data in
the write buffer can be expected in the A64FX, whereas the RIKEN simulator
is considered to have no write merge function. In addition, the reason
that the RIKEN simulator is nearly 10\% slower in cos and log 10 seems
to be the difference in gather load. In the A64FX, the combined gather
load allows up to 2 elements to be accessed in one cycle, while the
RIKEN simulator does not support combined gather load, and it takes 1
cycle per element. However, some other mathematical functions are
faster in the RIKEN simulator, and the reasons for this are now under
consideration.

\subsection{Evaluation of L2 cache and memory performance by multithreading}

Next, in order to evaluate the L2 cache and memory performance in
multithread execution, the performance of Stream Triad for two data
sizes was compared by changing the number of threads.

\begin{figure}[!t]
  \begin{center}
  \includegraphics[width=7cm]{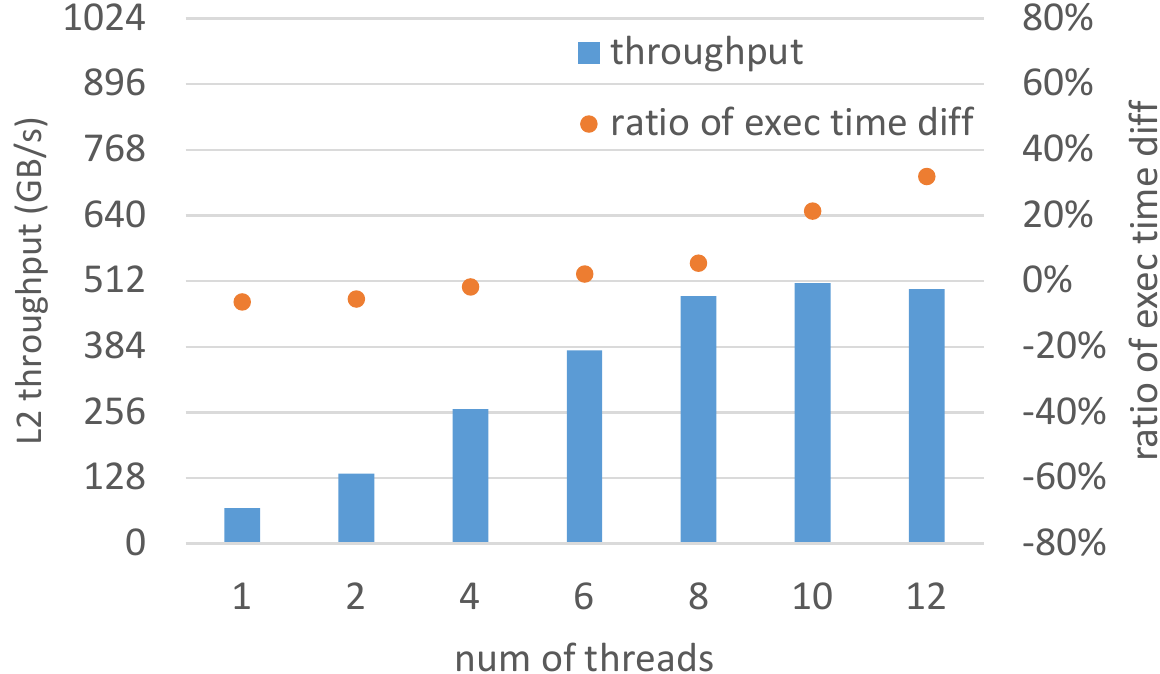}
  \caption{L2 cache throughput of Stream Triad in the RIKEN simulator and the difference in execution time with the test chip}\label{fig:streaml2}
  \end{center}
\end{figure}

The first evaluation is for L2 cache throughput, using Stream Triad for
the size within the L2 cache. The results are shown in Figure \ref{fig:streaml2}. These are the
optimized results when performing software prefetching four lines
ahead only for the L1 cache by compiler option. The bar graph is the
total L2 throughput in the RIKEN simulator when the number of threads
is changed from 1 to 12 for the same data size. The x-axis corresponds
to the number of threads, and the y-axis corresponds to the left
axis of the figure. Although the result is relatively scalable with the increase in
the number of threads, the throughput is saturated around 8
threads. The orange dots show the percentage difference in execution
time between the RIKEN simulator and the A64FX test chip. The y-axis
corresponds to the right axis of the figure. As with the kernel evaluation, a
positive difference in execution time indicates that the RIKEN
simulator execution time is larger than that of the A64FX test chip, and a negative
difference indicates that the A64FX test chip execution time is larger than that of
the RIKEN simulator. It can be seen from Figure \ref{fig:streaml2}
that when the number of threads is small, the RIKEN simulator is
somewhat faster, and when the number of threads is larger, the RIKEN
simulator becomes slower, and the difference between 10 and 12 threads
increases sharply and that the throughput at 12 threads is reduced by 30\% or more.

The reason why the RIKEN simulator is fast when the number of threads
is small is considered to be the difference in control of write back
from the L1 cache to the L2 cache. In the A64FX, write back is controlled exclusively with
other L1 cache accesses, but in the RIKEN simulator it is controlled
independently. On this point, we plan to make corrections so it will be the
same as the A64FX.

One of the reasons why the RIKEN simulator is slow when the number of
threads is large is considered to be the L2 cache fairness control. In
the A64FX, the L2 cache has a mechanism to fairly service requests from
each core, but this is not implemented in the RIKEN simulator. Therefore,
when the number of threads increases, performance variations among
threads occur, and the performance may be degraded because the
performance is pulled by the slowest thread. We are
also considering a mechanism to control some kind of fairness for the RIKEN simulator.

Moreover, the crossbar interconnect between the L1 cache and the L2 cache of
the A64FX is devised so that the performance degradation can be suppressed
even if the number of threads increases, and the performance is
improved to 12 threads in a scalable manner. On the other hand, in the
RIKEN simulator, crossbar performance is limited because a single
transmission source cannot transfer different transfer destinations
for transfer. Since it is considered difficult to correct the
performance limitation of this crossbar, it is under consideration
whether other adjustments can be made to match the performance.

\begin{figure}[!t]
  \begin{center}
  \includegraphics[width=7cm]{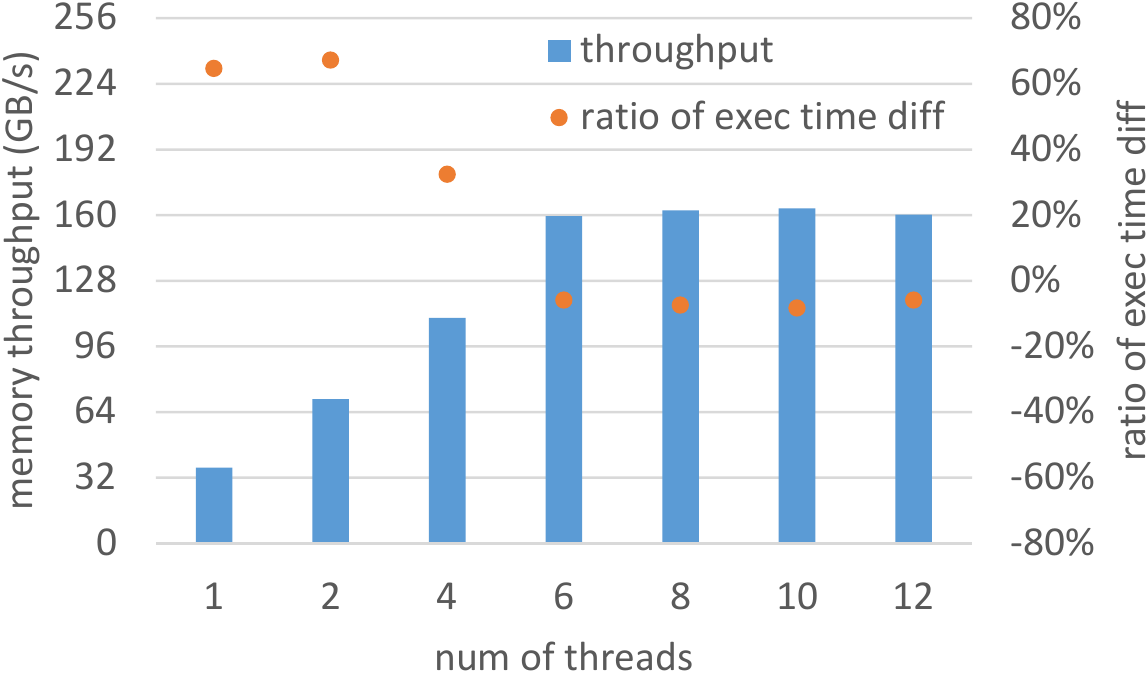}
  \caption{Memory throughput of Stream Triad in the RIKEN simulator and the difference in execution time with the test chip}\label{fig:streammem}
  \end{center}
\end{figure}

The second evaluation is for memory throughput, using Stream Triad for
twice the size of L2. The result is shown in Figure
\ref{fig:streammem}. The bar graph is the total memory throughput in
the RIKEN simulator when the number of threads is changed from 1 to 12
for the same data size. The x-axis corresponds to the number of
threads, and the y-axis corresponds to the left axis in the figure. It can be seen
that the memory throughput is saturated in six threads. The orange
dots show the percentage difference in execution time between the RIKEN
simulator and the A64FX test chip. The percentage of the difference
in execution time for more than 4 threads is 10\% or less. On the
other hand, when the number of threads is small, the percentage of the
difference exceeds 60\% and is quite large.

This is because the RIKEN simulator is not yet been implemented with the
hardware prefetch function for the L2 cache. Currently, a hardware prefetch
function for the L2 cache is being implemented and will be reevaluated after its
completion.

The memory throughput shown here is for a single CMG, and the total
memory throughput on the chip is 640 GB/s. This value is lower than
the memory bandwidth shown by \cite{A64FX-hotchip}, which is 830
GB/s. This is because the result of this evaluation is the memory
bandwidth seen from the application, and the throughput of the memory
itself exceeds 830 GB/s.  Also, while the result of this evaluation is
the memory bandwidth using hardware prefetch, the result of
\cite{A64FX-hotchip} is the result when applying software prefetch and
stream write optimization to the L2 cache. At present, these optimization
functions are being implemented in the RIKEN simulator, and will be
evaluated again after their completion.

\section{Conclusion}

RIKEN has developed a post-K processor simulator based on the
general-purpose processor simulator gem5. The simulator can estimate
the number of execution cycles accurately by simulating out-of-order
execution based on the basic architecture model with detailed
parameter tuning and function extension. This basic architecture model
is different in its details from the Fujitsu A64FX processor.
However, we aim to estimate the execution cycles of
applications on a single CMG of the A64FX with such an accuracy
that enables relative evaluations and tuning of applications.

The accuracy of this simulator was confirmed by comparing it with the
number of execution cycles of the A64FX test chip. For the
28 kernel programs, it was confirmed that the difference is 10\% or
less for 23 or 80\% or more of the kernel programs. In Stream Triad multi-threaded
execution, scalable performance according to the number of threads was
confirmed, but it was found that the difference in execution time is large
when the number of threads is large for L2 size data and the number of
threads is small for memory size data. The cause is that functions
such as target L2 prefetch are not implemented, and thus work on these
implementations will be continued.

The RIKEN environment where RIKEN simulators can be used was initially
provided only to users involved with key and emerging issues of the FS2020
project. However, since September 2018, RIKEN has been recruiting a wider
audience ~\cite{EVAL} through RIST (Advanced Information Technology
Research Organization). If you are interested in this simulator,
please consider using it.

\section*{Acknowledgments}
This work is partially funded by MEXT's program for the Development
and Improvement for the Next Generation Ultra High-Speed Computer
System, under its Subsidies for Operating the Specific Advanced Large
Research Facilities.


\begin{thebibliography}{1}

\bibitem{gem5}  
  N.~Binkert, B.~Beckmann, G.~Black, S.K.~Reinhardt, A.~Saidi, A.~Basu, J.~Hestness, D.R.~Hower, T.~Krishna, S.~Sardashti, R.~Sen, K.~Sewell, M.~Shoaib, N.~Vaish, M.D.~Hill and D.A.~Wood. \emph{The gem5 Simulator}, ACM SIGARCH Computer Architecture News, Vol.29, Issue 2, May 2011.

\bibitem{A64FX-hotchip}
  Y.~Yoshida, \emph{Fujitsu High Performance CPU for the Post-K Computer}, 2018 IEEE Hot Chips 30 Symposium, 2.13, Aug. 2018.

\bibitem{A64FX-cluster}
  Y.~Ajima, T.~Kawashima, Takayuki.~Okamoto, N.~Shida, K.~Hirai, T.~Shimizu, S.~Hiramoto, Yoshiro.~Ikeda, T.~Yoshikawa, K.~Uchida and T.~Inoue, \emph{The Tofu Interconnect D}, IEEE International Conference on Cluster Computing, pp.646-654, Sep. 2018.
  
\bibitem{SVE}
  N.~Stephens, S.~Biles, M.~Boettcher, J.~Eapen, M.~Eyole, G.~Gabrielli, M.~Horsnell, G.~Magklis, A.~Martinez, N.~Premillieu, A.~Reid, A.~Rico and P.~Walker, \emph{The ARM Scalable Vector Extension}, IEEE Micro, Vol.37, Issue 2, March/April 2017, pp.26--29.

\bibitem{K-computer}
  http://www.fujitsu.com/downloads/JP/archive/imgjp/jhpc/sparc64viiifx-extensions.pdf, \emph{SPARC64 VIIIfx Extensions}

\bibitem{EVAL}
  http://www.hpci-office.jp/pages/e\_other\_submission, \emph{Call for Proposals - Post-K Computer Performance Evaluation Environment}.
  
\end{thebibliography}
\end{document}